\documentclass[12pt,twoside]{article}
\usepackage[utf8]{inputenc}
\usepackage{amsmath}
\usepackage{amsfonts}
\usepackage{amssymb,bbold}
\usepackage{amsthm}
\usepackage{subfigure}
\usepackage{graphicx}
\usepackage{booktabs}
\usepackage{setspace}
\usepackage{wrapfig}
\usepackage{booktabs}
\usepackage{multirow}
\usepackage{float}
\usepackage{multicol}
\usepackage{url}
\usepackage{breqn}
\usepackage{longtable}
\usepackage{natbib}

\allowdisplaybreaks 
\usepackage{color}

\usepackage{arydshln}
\newcommand{\iid}{\overset{iid}{\sim}} 
\newcommand{\E}{\mathrm{E}}
\newcommand{\x}{\mathrm{x}}

\newcommand{\N}{\mbox{{\small\textsc{N}}}}

\newcommand\independent{\protect\mathpalette{\protect\independenT}{\perp}}
\def\independenT#1#2{\mathrel{\rlap{$#1#2$}\mkern2mu{#1#2}}}

\begin{document}

\title{Atlantic Causal Inference Conference (ACIC) Data Analysis Challenge 2017}
\author{P. Richard Hahn\footnote{Thanks to the Booth School of Business for financial support. Thanks also to Jinhui Xu, Yao Shi, and Changhan He for typesetting assistance. }, Vincent Dorie, and Jared S. Murray}
 \date{\today}

\maketitle
\begin{abstract}
This brief note documents the data generating processes used in the 2017 Data Analysis Challenge associated with the Atlantic Causal Inference Conference (ACIC). The focus of the challenge was estimation and inference for conditional average treatment effects (CATEs) in the presence of targeted selection, which leads to strong confounding. The associated data files and further plots can be found on the first author's web page.
\end{abstract}
\onehalfspacing
\section{Introduction}
The purpose of the Data Analysis Challenge is to provided a blinded empirical evaluation of causal inference methods on synthetic data, where the true data generating process is revealed after estimation and inference accuracy for each method has been determined. While any given simulation study is necessarily limited, here the separation between the data generators and the method submitters lends additional credibility to the results. The options for empirically evaluating causal methods (generating counterfactual predictions) are also more limited than for traditional (factual) prediction problems, where performance can be evaluated on test subsets of ``real'' data. The goal here is to design high-quality simulation studies, in the sense that the synthetic data are plausibly representative of real data in key respects.
%

\section{Overview}\label{overview}
For the 2017 Challenge, the data were generated according to 32 distinct, fixed, data generating processes (DGPs). For each of these, 250 independent replicate data sets were produced, for a total of 8,000 data sets. All 32 DGPs had covariate-dependent treatment effects.  We assumed that

\begin{equation}\label{mean}
\E(Y_i \mid X_i = \x_i, Z_i = z_i) = \mu(\x_i) + \tau(\x_i) z_i.
\end{equation}
From a structural equation vantage point, our response variable arises as 

$$Y_i = F(\x_i, z_i, \epsilon_i),$$
with treatment $Z_i$ arising as 
$$Z_i = G(\x_i, \nu_i)$$
for fixed, known functions $F$ and $G$. Exogenous noise variables were generated $\nu_i \independent \epsilon_i \independent Z_i$. In terms of potential outcomes we assume that

\begin{equation}
\E(Y^1 \mid \x) = \mu(\x) + \tau(\x),
\end{equation}
and
\begin{equation}
\E(Y^0 \mid \x) = \mu(\x),
\end{equation}
so that
\begin{equation}
\E(Y^1 - Y^0 \mid \x)  = \tau(\x)
\end{equation}
where $Y^1, Y^0$ are the potential outcomes under treatment and control, respectively, so that $\tau(\x)$ explicitly represents the CATE. Strong ignorability holds in every scenario considered here. Further, the stronger condition of no unmeasured treatment moderation holds -- that is, there are no unmeasured factors that drive treatment effect heterogeneity.

The 32 distinct DGPs were arrived at by considering four different types of errors:
\begin{itemize}
\item additive, independent, identically distributed,
\item additive, group correlated,
\item additive, heteroskedastic,
\item non-additive, independent, identically distributed.
\end{itemize}
The error terms were generated according to Gaussian distributions in all cases. 

Within each of these four error types, we considered ``high'' and ``low'' settings of three separate aspects of the DGP:
\begin{itemize}
\item magnitude of the causal effect,
\item strength of the confounding,
\item noise level in the response variable.
\end{itemize}
Interesting aspects of causal DGPs that were not considered this year include:
\begin{itemize}
\item non-Gaussian errors,
\item high dimensional covariates,
\item null effects.
\end{itemize}

\section{Control and/or moderating variables}
Across all 8,000 data sets, the measured control and moderating variables remain fixed, taken from the Infant Health and Development Program, or IHDP \citep{brooks1992effects}. These covariates were also used in the 2016 ACIC Data Analysis Challenge \citep{dorie2019automated}.  Only eight covariates out of 58 from original data were used:

\begin{itemize}\addtolength{\itemsep}{-0.5\baselineskip}
\item $X_{1}$: Mother's age (continuous),
\item $X_{3}$: Mother's cigarettes per day (continuous),
\item $X_{10}$: Mother's endocrine condition (binary),
\item $X_{14}$: Mother's nervous system condition (binary),
\item $X_{15}$: Mother's obstetric complications (binary),
\item $X_{21}$: Mother's birth place (categorical),
\item $X_{24}$: Mother's race (binary),
\item $X_{43}$: Child's bilirubin (continuous).
\end{itemize}
The correlations between these variables are shown in Table \ref{correlations} for reference. 
\begin{table}[!ht]
\centering
\begin{tabular}{r r r r r r r r r}
\toprule
correlation & $X_{1}$ & $ X_{3}$ & $X_{10}$ & $X_{14}$ & $X_{15}$ & $X_{21}$ & $X_{24}$ & $X_{43}$ \\
\midrule
$X_{1}$ & 1.00  & 0.04 & -0.07 & -0.03 & -0.04 & -0.07 & 0.03 & -0.01  \\
$X_{3}$ & 0.04 & 1.00 & -0.02 & 0.03 & -0.02 & -0.10 & -0.16 & 0.13 \\
$X_{10}$ & -0.07 & -0.02 & 1.00 & 0.04 & 0.09 & -0.02 & -0.10 & -0.07 \\
$X_{14}$ & -0.03 & 0.03 & 0.04 & 1.00 & 0.09 & -0.03 & -0.08 & 0.07 \\
$X_{15}$ & -0.04 & -0.02 & 0.09 & 0.09 & 1.00 & -0.03 & 0.04 & -0.04 \\
$X_{21}$ & -0.07 & -0.10 & -0.02 & -0.03 & -0.03 & 1.00 & 0.20 & -0.00 \\
$X_{24}$ & 0.03 & -0.16 & -0.10 & -0.08 & 0.04 & 0.20 & 1.00 & -0.11  \\
$X_{43}$ & -0.01 & 0.13 & -0.07 & 0.07 & -0.04 & -0.00 & -0.11 & 1.00  \\
\bottomrule
\end{tabular}
\caption{Correlation matrix of control variables.}\label{correlations}
\end{table}

\section{Data generation details}
\subsection{Additive errors}
We denote the varying settings by variables $\xi$, $\eta$, and $\kappa$. The variable $\xi$ modulates the magnitude of the effect size, and takes one of two values, 2 or 1/3. The variable $\eta$ modulates the standard deviation of the error, and takes one of two values, 5/4 or 1/4. The vector variable $\kappa$ corresponds to regression coefficients governing the propensity to receive treatment, and takes values $(3, \, -1)$ or $(0.5, \, 0)$. According to the values of $(\xi, \eta, \kappa)$, one obtains eight cases, as shown in Table \ref{cases}.

\begin{table}[!ht]
\centering
\begin{tabular}{rrrr}
\toprule
No. & Effect magnitude $\xi$ & Noise level $\eta$ & Selection strength $(\kappa_{1}, \, \kappa_{2})$ \\
\midrule
$1.$ & Low (1/3)  & Low (0.25) & Weak (0.5 , 0) \\
$2.$ & Low (1/3) & Low (0.25) & Strong (3 , -1) \\
$3.$ & Low (1/3) & High (1.25) & Weak (0.5 , 0) \\
$4.$ & Low (1/3) & High (1.25) & Strong (3 , -1) \\
$5.$ & High (2) & Low (0.25) & Weak (0.5 , 0) \\
$6.$ & High (2) & Low (0.25) & Strong (3 , -1) \\
$7.$ & High (2) & High (1.25) & Weak (0.5 , 0) \\
$8.$ & High (2) & High (1.25) & Strong (3 , -1)  \\
\bottomrule
\end{tabular}
\caption{Eight different cases and the values of corresponding parameters.}\label{cases}
\end{table}
\newpage
Next, define the following functions:
\begin{eqnarray*}
f(\x) &=& x_1+x_{43}+0.3(x_{10}-1), \\
 \pi(\x) &= &\mbox{Pr}(Z_i = 1) = (1+\exp{(\kappa_{1}f(\x)+\kappa_{2})})^{-1},\\
\mu(\x) &=& -\sin(\Phi(\pi(\x)))+x_{43}, \\
\tau(\x) &=& \xi(x_3 x_{24}+(x_{14}-1)-(x_{15}-1)), \\
\sigma(\x) &=& 0.4+\frac{x_{21}-1}{15},
\end{eqnarray*}
where $\Phi(\cdot)$ denotes the cumulative distribution function of a standard normal random variable. Let $\sigma_{y} = \eta\sqrt{\mathrm{Var}(\mu(\x)+\pi(\x)\tau(\x))}$, where the variance is taken over the observed sample. Finally, let  $\varepsilon_i \iid \N(0,1)$. With these definitions in hand, we can now describe the data generation protocols as follows.

\begin{itemize}
\item In the {\em independent, identically distributed errors} case we simply have
$$Y_i = \mu(\x_i)+\tau(\x_i)Z_i+ \sigma_y\varepsilon_i.$$

\item In the {\em group correlated errors} case we have
$$Y_i=\mu(\x_i)+\tau(\x_i)Z_i+\sigma_y(0.9\varepsilon_i + 0.1\epsilon_{x_{i,21}}),$$
where $\epsilon_{x_{i,21}}$ denotes one of sixteen realizations of a standard normal random variable, indexed by the variable $x_{21}$, which takes sixteen levels (recorded as 1 through 16). Accordingly, 10\% of the  error term is shared among variables with common values of $x_{21}$ (mother's place of birth).

\item In the {\em heteroskedastic error} case, we have
$$Y_i=\mu(\x_i)+\tau(\x_i)Z_i+\sigma(\x_i)\sigma_y\varepsilon_i,$$ 
so that the error variance varies as a function of $x_{21}$ (mother's place of birth). Observe that $\sigma(\x)$ ranges from 0.4 to 1.4 in constant increments. 
\end{itemize}
\subsection{Non-additive errors}
The notation for the non-additive case is considerably more cumbersome. Operationally, we simply generate data according to an additive error model and then perform a nonlinear transformation. However, in doing so we must correct for the causal effect, which is no longer explicitly defined, as in the additive case. Secondly, we endeavored to select the numerical constants governing the transformation so that the distribution of the outcome variable did not look dramatically different than it did in the additive cases. That is, the data is generated according to
\begin{eqnarray*}
\tilde{Y}_i &=& \tilde{\mu}(\x_i)+ \tilde{\tau}(\x_i)Z_i+\sigma_{y}\varepsilon_i, \\ 
Y_i &=& 13\Phi(\tilde{Y}_i \mid a, b) -6,
\end{eqnarray*}
where $\Phi(\cdot \mid a, b^2)$ denotes the CDF of a $\N(a,b^2)$ random variable and the functions are defined as above. Note that for $\tilde{\mu}(\x)$ and $\tilde{\tau}(\x)$ we use the definitions given above for $\mu(\x)$ and $\tau(\x)$; the new notation highlights that these functions no longer bear a direct relationship to the potential outcomes as described in Section \ref{overview}. Next, we determine specific values of $a$ and $b$ as:
\begin{eqnarray*}
a &=& \E(\tilde{Y}) = \frac{1}{n} \sum_i \tilde{\mu}(\x)+\tilde{\tau}(\x)\pi(\x)\\
b &=& 1.25 \sqrt{\mathrm{Var}(\tilde{Y})} = 1.25 \sqrt{\sigma^{2}_{y}+\mathrm{Var}(\tilde{\mu}(\x)+\tilde{\tau}(\x)\pi(\x)} 
\end{eqnarray*}
where the variance is taken over the sample. 

Finally, to compute the CATE, we employ the following fact concerning a standard normal random variable $W \sim \N(0,1)$:
$$\E(\Phi(m + sW)) = \Phi\left(\frac{m}{\sqrt{1 + s^2}}\right).$$
Applying this to our specific case, we write 
\begin{eqnarray*}
m^{1}(\x_i) &=& \frac{\tilde{\mu}(\x_i)+\tilde{\tau}(\x_i)-a}{b}, \\
m^{0}(\x_i) &=& \frac{\tilde{\mu}(\x_i)-a}{b}, \\
s &=& \sigma_y/b,
\end{eqnarray*}
and find that
\begin{eqnarray*}
\E(Y^{1} \mid \x) &=& 13\Phi\left(\frac{m^{1}(\x)}{\sqrt{\frac{\sigma^{2}_{y}}{b^2}+1}}\right)-6 ,\\
\E(Y^{0} \mid \x) &=& \mu(\x) \; = \;13\Phi\left(\frac{m^{0}(\x)}{\sqrt{\frac{\sigma^{2}_{y}}{b^2}+1}}\right)-6.
\end{eqnarray*}
It follows that the treatment effect can be calculated as
$$ \tau(\x) = \E(Y^{1} \mid \x)-E(Y^{0} \mid \x)= 13\Phi\left(\frac{m^1(\x)}{\sqrt{\frac{\sigma^{2}_{y}}{b^2}+1}}\right)-13\Phi\left(\frac{m^0(\x_i)}{\sqrt{\frac{\sigma^{2}_{y}}{b^2}+1}}\right). $$

\section{Targeted selection}
The main theme (unannounced) of the challenge this year is the notion of ``targeted selection''. That is, we were interested in understanding the behavior of various methods in situations where the likelihood that an individual receives treatment is a function of the expected response of that individual if left untreated. In the DGPs above this manifest as $\mu(\x)$ being a function of $\pi(\x)$; in retrospect it would have been more clearly experessed the other way around, with $\pi(\x)$ being written as a noisy function of $\mu(\x) = \E(Y^0 \mid \x)$. This type of confounding structure is highly plausible in many real world scenarios where the course of treatment is directly predicated on a prognosis determined by observed covariates, yet it is unlikely to arise from DGPs that are generated stochastically using polynomial bases (such as the 2016 Data Challenge). We hope that others are encouraged to investigate this type of DGP in future methodological development.

\section{Data files}
\subsection{File structure}
The exact data files used are available in the contest_data folder, stored with the following file structure. There are four folders, named
\begin{itemize}
\item {\tt group_corr}
\item {\tt heteroskedastic}
\item {\tt iid}
\item {\tt non-additive}.
\end{itemize}
Within each folder are eight subfolders, named according to a binary encoding of the simulation settings --- the first bit corresponds to the effect magnitude parameter $\xi$, the second bit corresponds to the noise level parameter $\eta$, and the third bit corresponds to the selection strength parameter vector $\kappa$. In all cases, a zero corresponds to the low setting and a 1 corresponds to the high setting. 

Within each of these subfolders there are 253 individual comma separated value files ({\tt .csv}). The files named {\tt 1.csv} through {\tt 250.csv} correspond to the 250 replicated data sets of the given DGP. The files named {\tt 1.***.y.csv} and {\tt 1.***.z.csv} are demo files that were provided to participants prior to submission; the response variable file was generated using a {\em different} set of treatment variables than are given in the treatment indicator file. Finally, the {\tt dgp.csv} file contains two columns, one of which contains the (conditional) treatment effect (somewhat confusingly referred to as ``alpha'' in the files, whereas ``tau'' would make better sense relative to this write-up) and the other which contains the corresponding mean of the untreated potential outcome ($\mu(\x)$).\\

\noindent {\color{red} NOTE: There is an error in the data files generated with heteroskedastic errors. The treatment effect for these files was, erroneously, calculated according to the formulas for the non-additive error case. We are leaving these files for documentation purposes, but they are incorrect. The results for the heteroskedastic errors are likewise incorrect and we omit them from further consideration.}

\subsection{Covariate matrices}
Common to all of these files is a shared covariate matrix, given as {\tt X.csv}. It contains $n = 4,302$ rows, corresponding to individuals in the original IHDP data set. This number is smaller by 500 than the covariate matrix used in the 2016 Data Analysis Challenge; 250 observations were provided to participants in advance and were used to generate the response variables given in the {\tt 1.***.y.csv} files and another 250 were used similarly to generate the {\tt 1.***.z.csv} files. These matrices are called {\tt X_subset_y.csv} and {\tt X_subset_z.csv} respectively. Additionally, the data generating processes described above actually utilized a transformed set of covariate values (the same as were used in the 2016 Data Analysis Challenge); these transformed variables are provided as {\tt Xt.csv}, {\tt Xt_subset_y.csv} and {\tt Xt_subset_z.csv}.   

\subsection{DGP code}
An {\tt R} script {\tt generate_data.R} is also provided, which implements the DGPs as described and generates the file structure describe above. At present, the code is not well formatted or documented. \\

\noindent {\color{red} NOTE: The code for generating the heteroskedastic and non-additive error data is correct, provided that the file names and flags are set by the user appropriately.}

\section{Evaluation: estimands and criteria}
Each submitted method was evaluated according to three criteria: root mean squared estimation error on the average treatment effect {\em on the treated} (ATT), coverage rate of interval estimates for the ATT (for nominal 95\% intervals), as well as root mean squared estimation error on the conditional average treatment effects (CATE), averaged over each observation in the sample, and average coverage of the CATEs. We also considered the length (respectively, the average length) of the reported intervals. Concretely, denote the conditional average treatment effect as
$$\tau(\x_i) =  \E(Y_i^1 \mid \x_i) - \E(Y_i^0 \mid \x_i) = \E(Y_i \mid \x_i) - \E(Y_i \mid \x_i)$$
where the last expression follows from strong-ignorability. Denote the average treatment effect on the treated as
$$\bar{\tau}_{att} = \frac{1}{n_t} \sum_{i, Z_i = 1} \lbrace \E(Y^1 \mid \x_i) - \E(Y^0 \mid \x_i) \rbrace = \frac{1}{n_t} \sum_{i, Z_i = 1} \tau(\x_i).$$

The root mean square estimation error for the CATE is computed via

$$\mbox{rmse}_{cate} = \sqrt{\frac{1}{n}\sum_i (\hat{\tau}(\x_i) - \tau(\x_i))^2}$$

and for the ATT it is computed as

$$\mbox{rmse}_{att} = \sqrt{(\hat{\tau}_{att} - \bar{\tau}_{att})^2}  = | \hat{\tau}_{att} - \bar{\tau}_{att}|.$$
 Note that in some publications $\mbox{rmse}_{cate}$ is called the ``precision in the estimated heterogeneous effects'' or PEHE. 
 
 Coverage is computed as
 $$\mbox{cover}_{att} = \frac{1}{m} \sum_j \mathbb{1}(l < \bar{\tau}_{att} < u)$$
where $l$ and $u$ are the lower and upper bounds of the reported interval estimate and the sum is taken over each replicate (specifically, $m = 250$ in this simulation). Likewise,
 $$\mbox{cover}_{cate} = \frac{1}{m} \sum_j \frac{1}{n} \sum_i \mathbb{1}(l(\x_i)i < \tau(\x_i) < u(x_i)).$$ Similarly, one can consider this quantity averaged only over the treated (respectively, untreated) units:

 $$\mbox{cover}_{catt} = \frac{1}{m} \sum_j \frac{1}{n_t} \sum_{i,Z_i = 1} \mathbb{1}(l(\x_i)i < \tau(\x_i) < u(x_i)).$$

An {\tt R} script implementing these functions is provided as {\tt evaluation_functions.R}.

\section{Teams and partial results}

Twenty-one entries were accepted and evaluated. Four of these entries were submitted by us on behalf of the ACIC conference. The first was a simple linear model for a straw man comparison. The second was a method utilizing the Bayesian additive regression tree (BART) model as implemented in the {\tt dbarts} package; this was a high-performing entry from the 2016 version of the challenge. The latter two methods, Bayesian Causal Forests and Gradient Random Forests, were submitted by us, {\em ex post}, in response to results that were presented at ACIC 2017.  The submission scripts for these two methods (21 and 22, below) are provided as {\tt bcf.att.R} and {\tt grf.att.R} respectively.

 We will not describe the submitted methods in detail here; we merely provide a suggestive entry name and the names of team members.
{\small
\begin{enumerate}\addtolength{\itemsep}{-0.5\baselineskip}
\item Linear model
\item BART
\item Super Learner + Target Maximum Likelihood Estimation (TMLE)
\item h20 Grid
\item Propensity score regression
\item BART (multiple chains)
\item BART (symmetrized)
\item BART with propensity score
\item BART with propensity score (symmetrized)
\item BART + TMLE
\item BART + inverse probability of treatment weighting (IPTW)
\item BART + inverse probability of treatment weighting (IPTW) (symetrized)
\item Targeted Learning
\item BART with influence function
\item Super Learner
\item Sparse regression
\item X-Learner BART
\item X-Learner hRF
\item Good Cause 1
\item Good Cause 2
\item Bayesian Causal Forest\footnote[1]{Submitted after preliminary results were reported at ACIC 2017.}
\item Gradient random forest\footnotemark[1]
\end{enumerate}
}
Note that method 3 does not estimate CATEs.

Because many authors submitted multiple methods, there were far fewer teams than there were methods. Using the numbering of methods from above:

\begin{itemize}\addtolength{\itemsep}{-0.5\baselineskip}
\item method 3: Susan Gruber and Mark van der Laan,
\item method 4: Frederico Nogueira,
\item method 5: Xu Shi,
\item methods 6 - 12: Nicole Bohme Carnegie and Jennifer Hill,
\item method 13: Chris Kennedy, Jonathan Levy, Caleb Miles, Ivana Malenica, Nima Hejazi, Andrew Kurepa Waschka and Alan Hubbard,
\item method 14: Razieh Nabi Abdolyousefi, Illya Shpitser, Razieh Nabi and Alex Gain,
\item method 15: Ryan Andrews, Illya Shpitser, Razieh Nabi and Alex Gain,
\item method 16: Marc Ratkovic,
\item methods 17 and 18: Soeren Kuenzel, Jasjeet Sekhon, Peter Bickel and Bin Yu,
\item methods 19 and 20: Naama Parush, Chen Yanover, Yishai Shimoni and Amit Gruber.
\item methods 1, 3, 21 and 22: P. Richard Hahn and Vincent Dorie 
\end{itemize}

\subsection{Summary plots}
Summary plots are supplied as a separate PDF file named {\tt eval_plots.R}. It contains results aggregated over the following subsets of the DGPs:

\begin{itemize}
\item all i.i.d. DGPs (8 total),
\item all non-additive error DGPs (8 total),
\item all group correlated error DPGs (8 total),
\item All homoskedastic DGPs (24 total) 
\end{itemize}
An {\tt R} script generating these plots is provided as {\tt summary_plots.R}; it pulls results data from {\tt .Rdata} files provided in the containing folder. 

\subsection{Summary of findings}

Because we considered a wide array of DGPs and several distinct evaluation criteria, it is difficult and also unhelpful to declare a ``winner''; the word ``challenge'', as opposed to ``contest'', is specifically intended to de-emphasize the competition aspect of this exercise. Still, certain notable trends did emerge, some of which we point out here.

\begin{itemize}
\item 95\% nominal coverage is achieved by no method. Despite the use of actual replicates, where only the noise variable changes, no method consistently covered at the desired rate. 
\item  Methods 8, 9 and 21 --- which specifically incorporate an estimate of the propensity score as covariates when estimating the response surface -- did particularly well in the targeted selection regimes studied this year. While there is a long literature establishing the utility of {\em combining} response surface and propensity score estimates for more efficient and robust estimation of aggregate treatment effects (e.g., \cite{bang2005doubly} and \cite{tmle}), to our knowledge most methods treat each prediction problem separately. Incorporating information about the selection process into response surface estimation seems worthy of further investigation; see \cite{bcf} for some initial investigations.
\end{itemize}

\bibliographystyle{abbrvnat} 
\bibliography{ACIC}

\end{document}